\documentclass[fleqn,usenatbib]{mnras}

\usepackage{float}
\usepackage{multicol}
\usepackage{graphicx}
\usepackage{subcaption}
\usepackage[T1]{fontenc}
\usepackage{hyperref}
\usepackage{ae,aecompl}

\usepackage{graphicx}	
\usepackage{amsmath}	
\usepackage{amssymb}	
\usepackage{url,color}
\usepackage{graphicx}
\usepackage{amsmath,float} 
\usepackage{amssymb}
\usepackage{multicol}
\newcommand{\ud}{\mathrm{d}}
\newcommand{\pd}{\partial}

\title[Differential Rotation in Shell Burning]{Differential Rotation in a 3D Simulation of Oxygen Shell Burning}

\author[McNeill and M\"uller]{
Lucy O.\ McNeill$^{1}$\thanks{E-mail: luc.mcneill@gmail.com }  and Bernhard M\"uller$^{1}$
\\
$^{1}$ School of Physics and Astronomy, Monash University, Victoria 3800, Australia\\
}

\date{Accepted XXX. Received YYY; in original form ZZZ}

\pubyear{2021}

\begin{document}
\label{firstpage}
\pagerange{\pageref{firstpage}--\pageref{lastpage}}
\maketitle

\begin{abstract}
We study differential rotation in late-stage shell convection in a 3D hydrodynamic simulation of a rapidly rotating $16M_\odot$ helium star with a particular focus on the convective oxygen shell. 
We find that the oxygen shell develops a quasi-stationary pattern of differential rotation that is neither described by uniform angular velocity as assumed in current stellar evolution models of supernova progenitors, nor by uniform specific angular momentum. Instead, the oxygen shell develops a positive angular velocity gradient with faster rotation at the equator than at the pole by tens of percent. We show that the angular momentum transport inside the convection zone is not adequately captured by a diffusive mixing-length flux proportional to the angular velocity or angular momentum gradient. Zonal flow averages reveal stable large-scale meridional flow and an entropy deficit near the equator that mirrors the patterns in the angular velocity. The structure of the flow is reminiscent of simulations of stellar surface convection zones and the differential rotation of the Sun, suggesting that similar effects are involved; future simulations will need to address in more detail how the interplay of buoyancy, inertial forces, and turbulent stresses shapes differential rotation during late-stage convection in massive stars. If convective regions develop positive angular velocity gradients, angular momentum could be shuffled out of the core region more efficiently, potentially making the formation of millisecond magnetars more difficult. Our findings have implications for neutron star birth spin periods and supernova explosion scenarios that involve rapid core rotation.
\end{abstract}

\begin{keywords}
convection -- hydrodynamics -- stars: interiors -- stars: massive -- stars: rotation
\end{keywords}


\section{Introduction}

Stellar rotation is a key factor in the evolution of massive stars \citep{Maeder2000,Langer2012}.  Reliable modelling of the rotational evolution of massive stars, which is determined by internal angular momentum transport, wind mass loss and binary interactions, is essential for understanding the spins of neutron stars and black holes, and the role of rotation in the supernova explosion mechanism. Accurate knowledge of the pre-supernova distribution of angular momentum is particularly important in the context of extremely energetic ``hypernovae'' and long gamma-ray bursts that may be produced by rapidly rotating ``millisecond magnetars'' \citep{Duncan1992,Thompson1993,Akiyama2003} or by black
hole accretion disks in the collapsar scenario \citep{Woosley1993,MacFadyen1999}. The problem of angular momentum transport in stellar interior is intimately connected with the theory of stellar dynamos \citep{brandenburg_05,charbonneau_14} and with the observed diversity of neutron star magnetic fields from
$10^{10}\texttt{-}10^{13}\, \mathrm{G}$  in normal pulsars to $\mathord{\sim} 10^{15}\, \mathrm{G}$ in observed magnetars, which yet
awaits an explanation \citep{Olausen2014,Kaspi2017}. Dynamo-generated fields will influence the rotational evolution of massive stars up to core collapse \citep{Spruit2002,Heger2005,Fuller2019}. Conversely, the  rotational evolution of massive stars will have a bearing on the magnetic fields of neutron stars, whether these are set by the pre-collapse field via flux conservation \citep{Ferrario2006} or generated after collapse by dynamo action \citep{Thompson1993}. 

Current stellar evolution models of supernova progenitors can treat rotation in a ``1.5D-approach'' that assumes shellular rotation and includes angular momentum transport processes by means of appropriate diffusion and/or advection terms
\citep{Heger2000,Meynet2000,Heger2005}\footnote{More sophisticated two-dimensional stellar evolution schemes promise a more rigorous treatment of stellar rotation
\citep{Deupree2000,Roxburgh2004,Rieutord2016}, but have not been applied to model the evolution of massive stars through advanced burning stages.}. State-of-the-art models \citep{Heger2000,Heger2005} account for a range of mixing processes, including the dynamical and secular shear instability \citep{Endal1978},
the Solberg-H{\o}iland instability \citep{Wasiutynski1946}, meridional circulation
\citep{Eddington1926,Kippenhahn1974,Zahn1992,Chaboyer1992}, the Goldreich-Schubert-Fricke instability
\citep{GS1967,Fricke1968}, and magnetic stresses generated by the Tayler-Spruit dynamo \citep{Spruit2002}. Within convective zones, the diffusion coefficient for angular moment is constructed from the turbulent velocity and correlation length, which can be estimated using mixing-length theory \citep[MLT,][]{Bohm1958}. It is usually assumed that the diffusive terms drive the angular momentum distribution towards uniform rotation, consistent with the notion that this represents the asymptotic state of thermodynamic equilibrium \citep{landau_5}.
Specifically, rigid rotation is established within convective regions within a few turnover times in current diffusion-based approximations for angular momentum transport.

There are abundant indications, however, that  current recipes for angular momentum transport in stellar evolution models require further refinement. For example, more efficient angular momentum transport is required to account for asteroseismic measurements  of low-mass stars \citep{Cantiello2014,Aerts2019}. What is most interesting among the observational findings in the context of this study, though, are indications that convective zones do \emph{not} evolve towards rigid rotation. Helioseismology reveals that the Sun's convective zone rotates 
$\mathord{\sim} 34 \%$ faster at the equator than near the poles ($470\, \mathrm{nHz}$ vs.\ $350\,\mathrm{nHz}$;  \citealp{Howe2009}).
Similar findings have been obtained for a few dozen solar-like stars via asteroseismology \citep{Benomar2018}, e.g., HD~173701 rotates twice as fast at the equator as near the pole.

The emergence of differential rotation is not entirely unexpected, and various ideas have been proposed to account for it. It has long been recognised that  meridional circulation, if properly modelled, can build up differential rotation \citep{Maeder1998,eggenberger_05}. In the context of the Sun, attempts have been made to explain the angular momentum distribution in the convection zone based on simple hydrodynamic principles by invoking thermal wind balance \citep{Kitchatinov1995,Balbus2009b}. Extending these ideas,  \citet{Jermyn2020a,Jermyn2020b} recently proposed  scaling laws for differential rotation in convective regions, which agree qualitatively with
global hydrodynamic and magnetohydrodynamic simulations
of stellar surface convection zones
\citep[e.g.,][]{Kapyla2011,Mabuchi2015} and rotation profiles determined via asteroseismology.   According to
\citet{Jermyn2020a,Jermyn2020b}, convective zones are expected to develop strong differential rotation in the case of high Rossby number (slow rotation) and asymptote to rigid rotation in the limit of small Rossby number (rapid rotation).

While  1D stellar modelling based on the mixing-length theory of convection \citep{Bohm1958}, with appropriate generalisations for the transport of composition and angular momentum by other instabilities \citep[e.g.,][]{Heger2000,Heger2005,Maeder2013} remain the state of the art for stellar evolution calculations over secular time scales, multi-dimensional simulations have opened a new window on mixing and angular momentum in stellar interiors.  There is already a long history of simulations that have addressed the rotation of the solar convection zone and the surface convection zones of other stars (e.g., \citealt{brun_04,matt_11,Kapyla2011,guerrero_13,Mabuchi2015,robinson_20}; and
\citealt{charbonneau_14} for a review). Despite a recent surge in in 3D simulations of late-stage convective burning
\citep{Meakin2007,BM2016,Jones2017,Cristini2017,MM2020,fields_20,yoshida_21,Yadav2020,Varma2021}, the interplay of convection and rotation has not yet been thoroughly explored by means of multi-dimensional hydrodynamic models  during the late convective burning stages, despite possible implications for the spins and magnetic fields of neutron stars and supernova explosion physics.
Due to the physical differences between the regimes of surface convection zone and late-stage convective burning (where radiative diffusion is completely unimportant), it is not clear whether the findings of the former can be translated directly to the interiors of massive stars. The problem of convection in rotating massive stars was only touched in passing by \citet{kuhlen_03} using anelastic 3D simulations with simplified source terms for nuclear burning. Subsequent
studies of convection in late-stage burning shells
of rotating progenitors by  \citet{Arnett2010} and \citet{chatzopoulos_16} were limited to axisymmetry (2D). Remarkably, \citet{Arnett2010} found that the convective shell adjusted  to constant specific angular momentum, which can be motivated by marginal stability analysis based on the Rayleigh criterion. This would be in stark contrast to the assumption of uniform rotation within convective zone in 1D stellar evolution models, and might seriously impact the secular evolution of the angular momentum profiles of massive stars on their way to core collapse. The assumption of axisymmetry, however, places a major caveat on these results. Recently, \citet{Yoshida2021} presented a short 3D simulation covering $91.6\,\mathrm{s}$ of oxygen shell burning in a rapidly rotating supernova progenitor, which showed adjustment towards constant specific angular momentum in the convective region. If confirmed, evolution towards constant specific angular momentum in convective zone might significantly alter the rotation profiles of massive stars compared to current 1D stellar evolution models, making it easier to maintain fast rotation in the stellar core region. This would be especially important in the context of progenitor scenarios for magnetorotational supernova explosions powered by ``millisecond magnetars'' \citep{Usov1992,Duncan1992,Akiyama2003}.
However, the short simulation time of \citet{Yoshida2021} calls for a closer look in longer simulations to confirm possible deviations from solid-body rotation in the inner convective zones of massive stars.
 
Here, we investigate the question of differential rotation in {rapidly rotating} convective regions {(i.e., with  rotational velocities exceeding convective velocities)} during oxygen burning for the first time in 3D, with a view to corroborating or improving stellar evolution models of massive stars, and with an eye to the possible implications for core-collapse supernovae and neutron stars. 
We perform a full $4 \pi$ solid angle simulation of
the shells around the iron core in a rapidly rotating
$16\, M_\odot$ helium star from a series of gamma-ray burst progenitor models by \citet{woosley_06}. We 
evolve the model for 12 minutes, or about 30 convective turnover times {in the oxygen burning shell}.  Using spherically averaged 1D profiles, an analysis of 2D zonal averages, and a spherical harmonics analysis, we attempt to better define the preferred rotation state in the convective region. We interpret our results in the light of proposed theories for differential rotation in convective zones and discuss implications for stellar evolution calculations.

Our paper is structured as follows: In Section \ref{section:stabilities}, we briefly review theoretical arguments about the angular momentum distribution in convective zones. In Section~\ref{sec:setup}, we describe the numerical methods and the setup of our 3D model. In Section \ref{section:results}, we analyse the differential rotation in our simulation using 1D and 2D flow decomposition, and discuss possible physical mechanisms for the unexpected angular momentum distribution, which neither evolves towards uniform specific angular momentum or uniform angular velocity. In Section~\ref{sec:conlclusions}{,} we summarise our findings and discuss possible implications for the evolution of massive stars and supernova scenarios involving rapid progenitor rotation.

\section{Brief Review of Arguments on the Rotational State of Convective Zones}

\label{section:stabilities}
Before discussing our simulation results, it is useful to briefly review key arguments that have hitherto been used to support or criticise the assumption of uniform rotation in {convective regions in} evolutionary models of massive stars.

\subsection{Angular Momentum Distribution in Convective Regions}
As the simplest argument for uniform rotation in convective zones, one might invoke the principle that only uniform rotation is allowed in
thermodynamic equilibrium \citep{landau_5}. For a given amount of angular momentum in a closed system, dissipation will ultimately drive the system towards uniform rotation as the state of minimal free rotational energy, which permits no further dissipation by viscous forces. This argument is specious, however, since stellar convection zones are open non-equilibrium systems, and must actually be continuously \emph{driven away} from thermodynamic equilibrium (by nuclear burning, neutrino cooling, or radiative/conductive heat transport to the surrounding shells) to maintain steady-state convection. Such systems will often settle into a quasi-equilibrium state, but this quasi-equilibrium is determined by dynamical considerations and does not necessarily resemble the thermodynamic equilibrium state, {e.g.}\ the non-isothermal stratification of convection zone and excited eccentricity in tidally interacting binaries \citep{mmm20}.

It has been argued that uniform rotation should indeed arise as such a dynamical quasi-equilibrium. The key idea is that horizontal shear is generally quickly eliminated by efficient turbulent angular momentum transport to enforce ``shellular rotation'' \citep{Zahn1992}, and that in convective zones efficient angular momentum transport by dynamical shear instability \citep{Zahn1974}
also eliminates vertical shear, so that the convective zone adjusts to a marginally stable state in which shear does not drive any further angular momentum transport on dynamical time scales. Essentially this argument already goes back to \citet{Endal1978}, who appealed to the effective turbulent viscosity provided by convective motions to enforce a state of marginal stability.

These arguments also leave loopholes, however. The assumption that the Richardson criterion for shear mixing can be adapted from planar flow in the form proposed by \citet{Zahn1974} is not above criticism because it is known that the centrifugal force and Coriolis force affect the stability of rotating flows. For isentropic, inviscid, cylindrical Taylor-Couette flow, linear stability is determined by the \emph{angular momentum} gradient according to the Rayleigh criterion \citep{Rayleigh1916}, i.e., an angular velocity profile that declines as $\Omega_z \propto R^{-2}$ with the distance $R$ from the rotation axis is marginally stable. The Rayleigh criterion can be generalised to the 
Solberg-H{\o}iland criteria to account for entropy and composition gradients and gradients parallel to the rotation axis
\citep{Solberg1936,Wasiutynski1946,Zahn1974,Tassoul1978}. One can argue that the
Solberg-H{\o}iland criteria are strictest
in the equatorial plane (assuming shellular rotation), and effectively reduce to the
stability criterion for the radial derivative of the pressure $P$, density $\rho$, and angular velocity $\Omega_z$, \citep{Endal1978,Heger2000}
\begin{equation}
-\frac{g}{\rho} \left[ \left(\frac{\pd \rho }{\pd r } \right) -
 \frac{\rho}{P} \frac{1}{\Gamma}\left(\frac{\pd P }{\pd r } \right)  \right]
 +
  \frac{1}{r^3} \frac{\partial r^4\Omega_z ^2}{\partial r} 
 \geq 0,
\end{equation}
where $g$ is the (effective) gravitational acceleration and $\Gamma$ is the adiabatic index. If convection flattens entropy and composition gradients, then marginal stability considerations would suggest that the shell-averaged angular momentum around
the rotation axis
$\langle l_z\rangle \sim r^2 \Omega_z$
is constant. In fact, one could go one step further and postulate that thorough mixing within a convective zone could homogenise the angular momentum $\mathbf{l}$ altogether since
$\mathbf{l}$ is conserved along pathlines in the absence of non-radial pressure gradients.
A spherically-averaged rotation profile
with $\langle l_z\rangle \sim r^2 \Omega_z=\mathrm{const.}$ has indeed been found in the 2D simulation of \citet{Arnett2010}, and 
\citet{Yoshida2021} also saw indications of adjustment towards
such a state in their recent 3D simulation.

These arguments still oversimplify the situation in convective zones, though. One obvious problem lies in the implication that
$\Omega_z$ should diverge at the origin in case of a core convection zone, but this is not the only problem. Even if the short-term effect of convective overturn may be to homogenise all advected quantities, i.e., entropy, mass fractions, and the specific angular momentum, fully developed quasi-stationary convection is generally characterised by a state of force balance or flux balance rather than marginal stability. For non-rotating convection this amounts to balance between energy gain or loss by heating and cooling, buoyant driving (which requires non-vanishing entropy contrasts) and turbulent dissipation \citep[e.g.,][]{Murphy2011}. In terms of the momentum budget, non-rotating steady-state convection at low-Mach number  are characterised by balance between the buoyancy forces and (turbulent) viscosity in a time-averaged sense, and the balance will result in a slightly superadiabatic gradient and the familiar scaling laws for the velocity and energy flux from mixing-length theory \citep{biermann_32,Bohm1958}. In the case of rotation, inertial forces need to be taken into in such balance consideration. In the inertial frame, balance between inertial forces due to rotation, pressure force, and gravity can be expressed in cylindrical coordinates $(R,\varphi,z)$ as
\begin{equation}
\label{eq:inertial_balance}
    \nabla P
    =
    -\rho \mathbf{g}
    +\rho R \Omega_z^2(R,z) \mathbf{e}_R,
\end{equation}
where $\mathbf{g}$ is the gravitational acceleration. After neglecting
horizontal pressure gradients, this can be recast as the thermal wind equation \citep[{e.g.}][]{Kitchatinov1995,Thompson2003,Balbus2009c}\footnote{Note that the thermal wind equation is usually derived
from the vorticity equation, but follows directly
from Equation~(\ref{eq:inertial_balance}) after assuming
a radial pressure gradient, which makes the concept of force
balance more obvious.},
\begin{equation}
\frac{\partial \Omega_z^2}{\partial r} - \frac{\tan \theta}{r} \frac{\partial \Omega_z^2}{\partial \theta} 
=
\frac{1}{\Gamma \rho r^2 \cos \theta\, \sin\theta }
\frac{\pd P}{\pd r}
\frac{\pd \sigma''}{\pd \theta},
\label{eq:TWE}
\end{equation}
where $\sigma''$ is the entropy contrast with respect to some purely radial reference entropy $\sigma(r)$ close to (but not necessarily identical to) the angle-averaged entropy.
The ``baroclinic'' term $\pd \sigma''/\pd \theta$ is
essentially a reflection of latitudinal variations in the
buoyancy force. \citet{Balbus2009a,Balbus2009b,Balbus2009c}
argued that together with a functional relationship
$\sigma''=\sigma''(\Omega_z^2)$ (which can be motivated
by various arguments, such as efficient mixing along surfaces
of constant $\Omega_z$), the thermal wind equation naturally
explains the solar rotation profile with faster rotation
at the equator. 

In addition to the inertial and buoyancy forces determined
by the large-scale, temporally and spatially averaged flow,
turbulent stresses can also affect the force balance and destroy
 uniform rotation by ``anti-diffusive'' transport of angular momentum by the $\Lambda$-effect \citep{Krause1974,Ruediger1989, Kichatinov1993}. In reality, both effects may play a role,
 and their relative contribution is still under debate in the case
 of the Sun \citep[e.g.,][]{Brun2010}.

\subsection{Relation to 1D Mixing Length Theory}
\label{section:MLT}
For use in spherically symmetric stellar evolution models, these (competing) considerations need to {be} approximated by some 1D conservation law for the specific angular momentum $\mathbf{l}$ in the spirit of MLT \citep{biermann_32,Bohm1958}.
Assuming shellular rotation about the same axis everywhere inside the star, this implies an equation for the angular velocity $\Omega_z$
about this axis,
\begin{equation}
\label{eq:mlt_omega}
\frac{\pd \rho r^2 \Omega_z}{\pd t}
+\frac{1}{r^2}\frac{\pd \rho v_r \Omega_z r^4}{\pd r}
+\frac{1}{r^2}\frac{\pd r^2 F_l}{\pd r}=0,
\end{equation}
with some flux function $F_l$ and an additional advective flux term
that accounts for spherical expansion or contraction.
Formally the choice of the flux function can be understood as a closure relation for a spherical Favre  decomposition of the flow (\citealp{Favre1965}; {see}\ Section~\ref{sec:1d_favre}), but implementations of angular momentum transport in 1D stellar evolution codes have historically often just exploited heuristic analogies for the diffusion approach to compositional mixing \citep{Eggleton1972}. This phenomenological approach is sufficient to understand the relation between the flux function and the corresponding steady-state angular momentum distribution in convective zones. 
If the diffusive flux is computed from the gradient of
the angular velocity \citep{Heger2000,Heger2005},
\begin{equation}
\label{eq:mlt_omegaz}
    \frac{\pd \rho r^2 \Omega_z}{\pd t}+
\frac{1}{r^2}
    \frac{\pd}{\pd r}
    \left( 
    \rho v_r \Omega_z r^4
    \right)
    =
    \frac{1}{r^2}
    \frac{\pd }{\pd r}
    \left(\rho D r^4 \frac{\pd \Omega_z}{\pd r}\right),
\end{equation}
with some appropriate MLT diffusion coefficient $D$, 
convective zones will be driven towards uniform rotation. If
the diffusive angular momentum flux is computed from the
specific angular momentum gradient instead,
\begin{equation}
\label{eq:mlt_l}
    \frac{\pd \rho r^2 \Omega_z}{\pd t}+
\frac{1}{r^2}
    \frac{\pd}{\pd r}
    \left( 
    \rho v_r \Omega_z r^4
    \right)
    =
    \frac{1}{r^2}
    \frac{\pd }{\pd r}
    \left(\rho D r^2 \frac{\pd \Omega_z r^2}{\pd r}\right),
\end{equation}
convective zones will adjust to $l_z=\mathrm{const}.$ instead.
This formulation might appear more in spirit with MLT, since the MLT energy and compositional fluxes are determined by the gradients of quantities that are conserved under advection, i.e., entropy and mass fractions. However, the situation is more subtle for angular momentum transport. By actually performing a proper turbulent flow decomposition under the assumption of shellular rotation, one instead obtains an advection-diffusion equation with an extra meridional circulation term that depends on the quadrupole component $v_\mathrm{Q}$
of the radial velocity \citep{Maeder1998},
\begin{equation}
    \frac{\pd \rho r^2 \Omega_z}{\pd t}+
    \frac{1}{r^2}
    \frac{\pd}{\pd r}
    \left( 
    \rho v_r \Omega_z r^4
    \right)
    =
    \frac{1}{r^2}
    \frac{\pd}{\pd r}
    \left[
    \rho  r^4
    \left(\frac{1}{5}
    {\Omega_z} v_\mathrm{Q}
     +
    D  \frac{\pd \Omega_z}{\pd r}
    \right)
    \right]
    \label{eq:MZ98}.
\end{equation}
The extra term arises because a first-order term in the Favre fluctuations survives in the spherically averaged
angular momentum equations (see Appendix~\ref{sec:appendix} for details). Although little has been made of the  meridional circulation term in the formulation of \citet{Maeder1998} in the context of differential rotation \emph{within} convection zones, it
is clear that it will force differential rotation if
the meridional circulation velocity $v_\mathrm{Q}$ is of the same
order as the convective velocity $v_\mathrm{conv}$, which 
determines the diffusion coefficient $D\sim v_\mathrm{conv} r$. In essence, the term is a 1D analogue to the $\Lambda$-term
from stellar turbulence theory \citep{Krause1974,Ruediger1989, Kichatinov1993}.

In the subsequent discussion, we shall evaluate these considerations in the context of late-stage convection in massive stars using our 3D hydrodynamics simulation.

\section{Shell Burning Simulation}
\label{sec:setup}
We have performed a 3D simulation of convective shell burning of model HE16O from \citet{woosley_06}. This model is a rotating $16 M_\odot$ helium star with a rotational velocity that is
of order of a few percent of the Keplerian velocity
in the oxygen shell, and similar to the convective
velocities, i.e., the convective Rossby number
is of order unity. The stellar evolution model
assumes shellular rotation and includes angular momentum transport by hydrodynamic processes and magnetic torques following \citet{Heger2000,Heger2005}.

\begin{figure}
  \centering
    \includegraphics[width=0.5\textwidth]{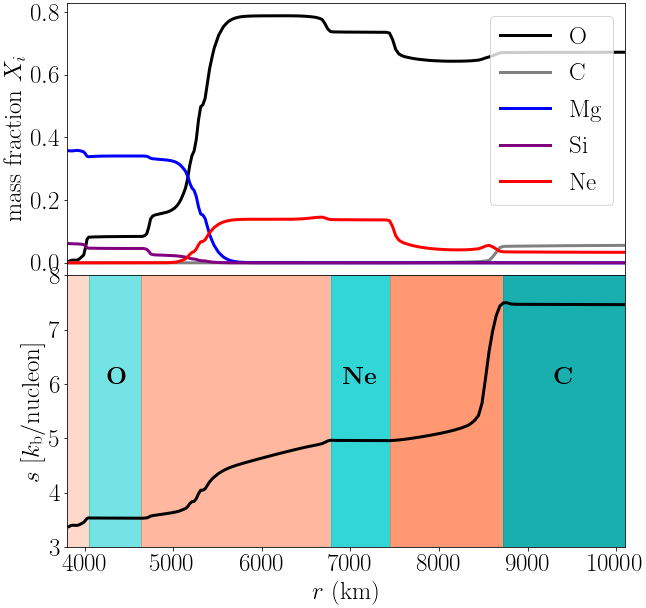}
  \caption{Initial chemical composition (top) and entropy profile (bottom) for the shell burning simulation. The model covers the region from $3,800\,\mathrm{km}$ to $31,200\, \mathrm{km}$ in radius, or $1.88\,M_\odot$ to $5.03\,M_\odot$
  in mass coordinate. There are three distinct convective burning regions (turquoise) which persist throughout the course of the simulation. From left to right, these are the oxygen {(1.95 to 2.10$M_\odot$)}, neon {(2.47 to 2.55$M_\odot$)}, and carbon {(2.71 to 5.03$M_\odot$)} burning shell, respectively. Convectively stable zones 
  (orange) correspond to regions
  with positive entropy gradients.}
\label{fig:composition}
\end{figure}

Following the methodology of \citet{BM2016}, we map the 1D stellar evolution model into the finite-volume code \textsc{Prometheus} \citep{fryxell_89,fryxell_91} at a late pre-collapse stage about 12 minutes before the onset of collapse. We then follow the evolution
to the onset of collapse, contracting the inner boundary following the trajectory of the corresponding mass shell in the stellar evolution model.
Our simulation includes the mass shells initially
located between $3,800\,\mathrm{km}$ and $31,200\, \mathrm{km}$, which covers
three distinct convective burning shells{. These are shown as  shaded cyan regions} in the progenitor entropy profile in Figure~\ref{fig:composition}, indicated by the three flat entropy gradients located around $4,200$, $7,000$ and $9,000\,\mathrm{km}$, corresponding to
oxygen burning, neon burning, and carbon burning, respectively. We shall mostly focus on the oxygen shell, which undergoes
about 13 turnover  over the period of interest (and about 26 for the whole simulation) to reach some form of quasi-steady state.

The extent and location of the convective and stable regions change somewhat over the course of the simulation because of the contraction of the shells, variations in burning rate, and convective entrainment.
The evolution becomes very dynamical in the last minutes prior to collapse with considerable modifications to the structure. Convection is fully developed by $100\, \mathrm{s}$ (corresponding to about three convective turnover times {for the oxygen shell}), after which time the inner convective region expands, then after $300\, \mathrm{s}$
 moves inwards again, staying roughly constant in radial extent for several minutes. 
 By $600\, \mathrm{s}$ the inner convectively stable region has disappeared completely, and the inner convective region covers around $1,000\,\mathrm{km}$,
 effectively doubling its width
 and growing by a factor three in mass by
 the time of core collapse due
 to a dramatic acceleration of entrainment right before collapse .
 Between about $300\, \mathrm{s}$ and $500 \, \mathrm{s}$, we have stable, quasi-stationary convection that is neither affected by initial transients, nor by the accelerating contraction right before collapse. Our analysis of differential rotation therefore mostly focuses on this phase of the simulation.

The numerical treatment is largely the same as in \citet{BM2016}. The simulation uses an overset Yin-Yang grid \citep{kageyama_04,melson_15a} with a resolution
of $450\times 56 \times 148$ zones in radius, latitude, and longitude on each Yin-Yang patch, corresponding to a resolution of $2^\mathrm{\circ}$ in angle. The momentum equation is reformulated according
to Equations~(23{,} 24) of \citet{Mueller2020} to improve angular momentum conservation. The 19-species nuclear reaction network of \citet{weaver_78} is used for nuclear burning.

\section{Results}
\label{section:results}
To analyse the simulation, we use Reynolds/Favre decomposition
\citep{Favre1965}  of the turbulent flow based
on azimuthal and spherical averages. For the spherical flow decomposition, we use hats (or, alternatively, angled brackets) and primes for the volume-weighted Reynolds averages and fluctuating components of volume-specific quantities such as the density $\rho$ and pressure $P$ (e.g., $\hat{\rho}$ and $\rho'$). These are defined for any such quantity $X$ as
\begin{eqnarray}
 \hat{X}(r) &=& \langle  X \rangle =  \frac{1}{4\pi}\int X \, \ud \omega,
 \label{eq:Favre1}
\\
 X'(r,\theta,\varphi)&=&X-\hat{X}(r),
 \label{eq:Favre3}
\end{eqnarray}
where we have used $\ud \omega=\sin \theta\,\ud\theta \,\ud\varphi$ instead of $\ud \Omega$ for the surface element on the unit sphere to avoid confusion with the angular velocity.
Mass-weighted Favre averages and fluctuating components of
mass-specific quantities like the internal energy density $\epsilon$
are denoted by tildes and double primes, e.g., $\tilde{\epsilon}$, and
$\epsilon''$. For any such quantity $Y$, we have
\begin{eqnarray}
 \tilde{Y} (r) &=& \frac{ \int \rho Y 
\, \ud \omega}{\int \rho
\, \ud \omega},
\\
Y''(r,\theta,\varphi)&=&Y-\tilde{Y}(r).
\label{eq:Favre2}
\end{eqnarray}
Averages of mass-specific  quantities
denoted by angled brackets are to be implicitly understood as mass-weighted averages, e.g., $\langle v_r''^2\rangle
=\langle \rho v_r''^2\rangle/\langle \rho\rangle$.
An additional complication arises for the angular velocity. In order to make the Favre average
$\tilde{\Omega}_z$ consistent with the Favre average
$\tilde{l}_z$ of the  angular momentum, we define
\begin{equation}
   \tilde{\Omega}_z=
   \frac{\int \rho \Omega_z r^2 \sin^2\theta \,\ud \omega}{\int \rho r^2 \sin^2\theta\,\ud \omega }=\frac{\tilde{l}_z}{\tilde{i}_{zz}},
   \label{eq:Omzav}
\end{equation}
where $\tilde{i}_{zz}$ is the $zz$-component of the gyration tensor of a mass shell,
\begin{equation}
    \tilde{i}_{zz}=
    \frac{\int \rho r^2 \sin^2\theta\, \ud \omega}{\int \rho \,\ud \omega}.
\end{equation}

For rotating flow,  a more fine-grained characterisation of the turbulent flow field is called for. It is useful to retain the dependence on $r$ and $\theta$ and only average in the $\varphi$-direction. We therefore calculate  mass-weighted azimuthal Favre averages for {any such} quantity $Z$ and denote them by bars (e.g., $\bar{\Omega}$) 
\begin{eqnarray}
     \bar{Z} (r,\theta) &=& \frac{ \int \rho Z
\, d \varphi}{\int \rho
\, d \varphi}.
\label{eq:angleav}
\end{eqnarray}

It is also useful to consider the azimuthal averages
of fluctuations with respect to the \emph{spherically}
averaged background flow, which we denote by triple primes, e.g., for quantity $Z$,  $Z''' (r,\theta)$ is defined as
\begin{equation}
         Z''' (r,\theta) = \langle Z(r,\theta,\varphi) - \tilde{Z}(r,\theta,\varphi)\rangle_\varphi=\frac{ \int \rho Z''
\, d \varphi}{\int \rho
\, d \varphi}.
\end{equation}
In other words, we take a density-weighted azimuthal average (denoted by angle brackets with index $\varphi$)
of the 3D fluctuations, which are calculated from the spherical average. This  is equivalent to subtracting the spherically averaged mean from the azimuthally averaged mean value of quantity $Z$.

\begin{figure}
  \centering
    \includegraphics[width=0.5\textwidth]{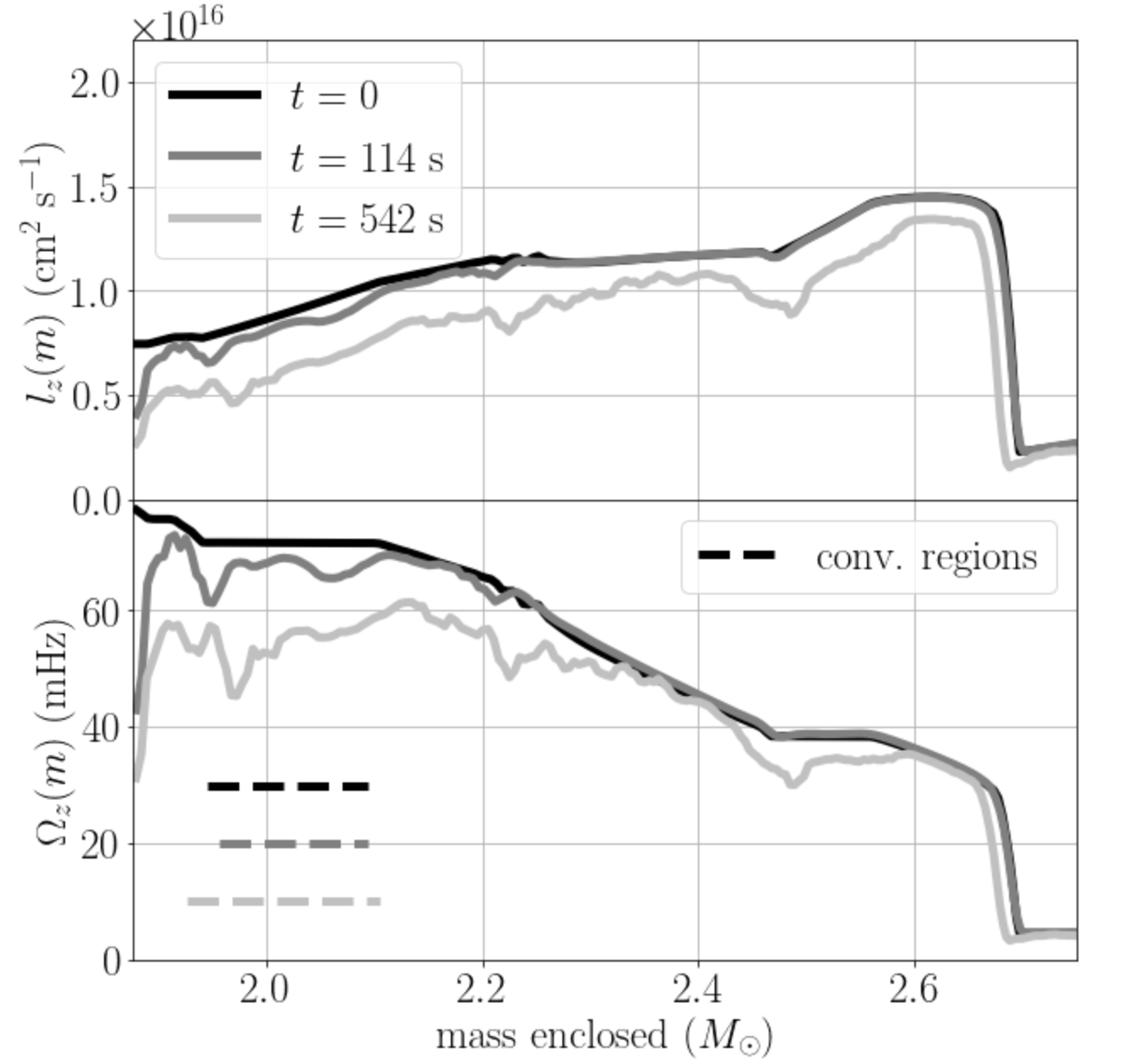}
  \caption{Spherically averaged {(mean)} specific angular momentum ${\langle}l_z{\rangle}$ (top panel) and angular velocity ${\langle}\Omega_z{\rangle}$ (bottom panel) as a function of enclosed mass $m$ at $0\, \mathrm{s}$ (black), $112\, \mathrm{s}$ (grey), and $542\, \mathrm{s}$ (light grey).
  The plots completely cover the inner two convective zones  around $1.95 \texttt{-} 2.10 M_\odot$
  and $2.45\texttt{-} 2.55M_\odot$, which are visible as flat plateaus in $\Omega_z$ at  the beginning of the simulation (black curve).
  The radial extent of the inner (oxygen) burning shell at these three times is indicated by dashed lines. Note that
  during the simulations $\mathord{\sim} 16 \%$ of angular momentum is lost by numerical conservation errors, which can be inferred from the changing integral under the curves in the top panel.
    As the simulation progresses, a positive
    angular velocity gradient emerges in the inner convective shell and the angular momentum gradient \emph{steepens}, indicating transport of angular momentum to larger radii. 
  }
\label{fig:AM-gradient}
\end{figure}

\subsection{Spherical Favre Analysis}
\label{sec:1d_favre}
 In Figure~\ref{fig:AM-gradient} we show the spherically averaged specific angular momentum 
$\tilde{l}_z$
 and angular velocity 
 $\tilde{\Omega}_z$
 as functions of mass coordinate $m$ (rather than $r$ to avoid changes in the location of convective zones for a better comparison by eye). The comparison of the initial rotation profile with rigidly rotating convection zones with profiles at $t=114\,\mathrm{s}$ (once convection is fully developed)  and $t=542\, \mathrm{s}$ (two minutes before collapse) reveals two important phenomena. The secular violation of angular momentum conservation implies that we must exercise caution in taking the emerging angular momentum distribution as a steady-state solution at face value. The differential rotation within the convective zones may still represent some form of quasi-steady state if it emerges on time scales shorter than the time scale for numerical loss of angular momentum. With this caveat, we find that a departure from rigid rotation develops during the simulation. Interestingly, a  (small) positive angular velocity
 gradient $\ud \tilde{\Omega}_z/\ud r$ develops, and the specific angular momentum is not mixed homogeneously. In fact, the angular momentum gradient become steeper than for rigid rotation, which is diametrically opposite to the emergence of a profile with constant specific angular momentum in the 2D model of \citet{Arnett2010}
 and the 3D simulation of \citet{Yoshida2021}. 

One might be concerned that angular momentum conservation errors have contributed to the development of this positive angular velocity gradient. However, comparing the turbulent 
angular momentum flux from the Favre analysis to the expected angular momentum flux from MLT provides more evidence that convection does not drive the angular velocity profile towards constant $\Omega_z$ (as assumed in current evolutionary models of massive stars) or constant $l_z$.

We first investigate the MLT flux that would be required to establish constant specific angular momentum throughout the convection zone.
The MLT angular momentum flux can be obtained 
from Equation~(\ref{eq:mlt_l}) by inserting the familiar expression for the diffusion coefficient
$D$ in terms of the convective velocity
$v_\mathrm{conv}$ and pressure scale height
$H_P$ \citep{weaver_78,Heger2000,BM2016},
\begin{equation}
    D=\alpha \, \alpha_3 H_P v_\mathrm{conv}.
\end{equation}
where $\alpha$ gives the ratio of the mixing length to the pressure scale height. The second dimensionless scaling factor $\alpha_3$ for the flux is commonly taken as $\alpha_3=1/3$ \citep{weaver_78,Heger2000}. For the sake
of simplicity, one can roll those two coefficients into one coefficient
$\alpha_3$ and implicitly set $\alpha=1$.
Expressing $v_\mathrm{conv}$ in terms of the radial
velocity fluctuations as $v_\mathrm{conv}=
\langle v_r''^2\rangle^{1/2} = \left( \langle\rho v_r''^2\rangle/\langle\rho\rangle\right) ^{1/2}$,
we find the
MLT flux $F_{\mathrm{MLT},l}$,
\begin{equation}
F_{\mathrm{MLT},l}=
4 \pi r^2 
\alpha\, \alpha_3
\rho \langle v_r''^2 \rangle^{1/2} H_P  \frac{\partial l_z}{\partial r},
\label{eq:MLTvr}
\end{equation}
which can be compared to the turbulent angular momentum flux from the Favre decomposition of the
angular momentum equation,
\begin{equation}
\frac{\partial \langle \rho l_z \rangle}{\partial t}
+
\nabla\cdot \mathbf{F}_\mathrm{adv}
+
\nabla\cdot \mathbf{F}_\mathrm{turb}
=0,
\end{equation}
with advective and turbulent fluxes
$\mathbf{F}_\mathrm{adv}$ and $\mathbf{F}_\mathrm{turb}$.
If $l_z$ were truly homogenised radially \emph{and} on shells, the background state of the Favre decomposition is $l_z=\mathrm{const.}$ on shells, and
we obtain,
\begin{equation}
\label{eq:fturbl}
F_{\mathrm{turb}} = {4 \pi r^2} \langle \rho v_r'' l_z''\rangle.
\end{equation}

\begin{figure*}
\begin{multicols}{2}
\includegraphics[width=\linewidth]{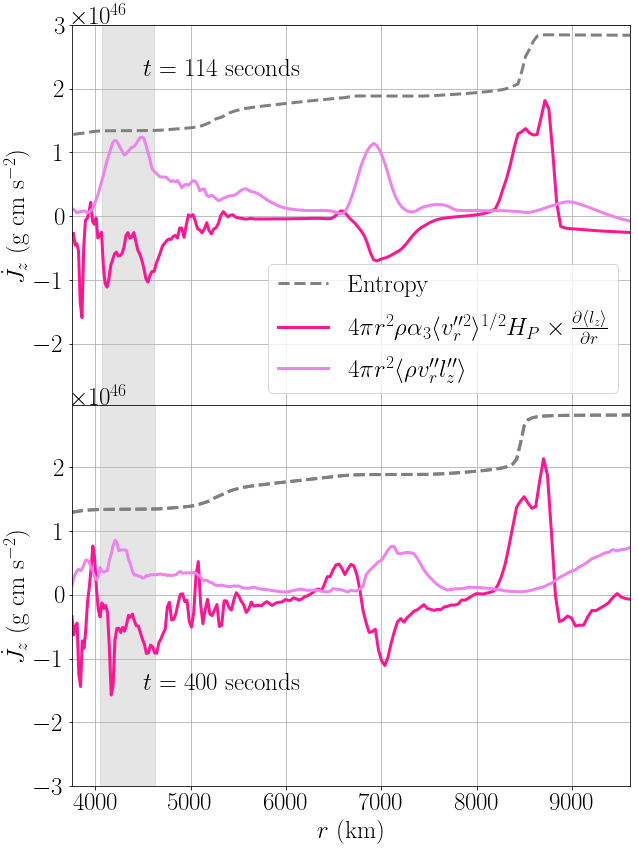}\par 
  \includegraphics[width=\linewidth]{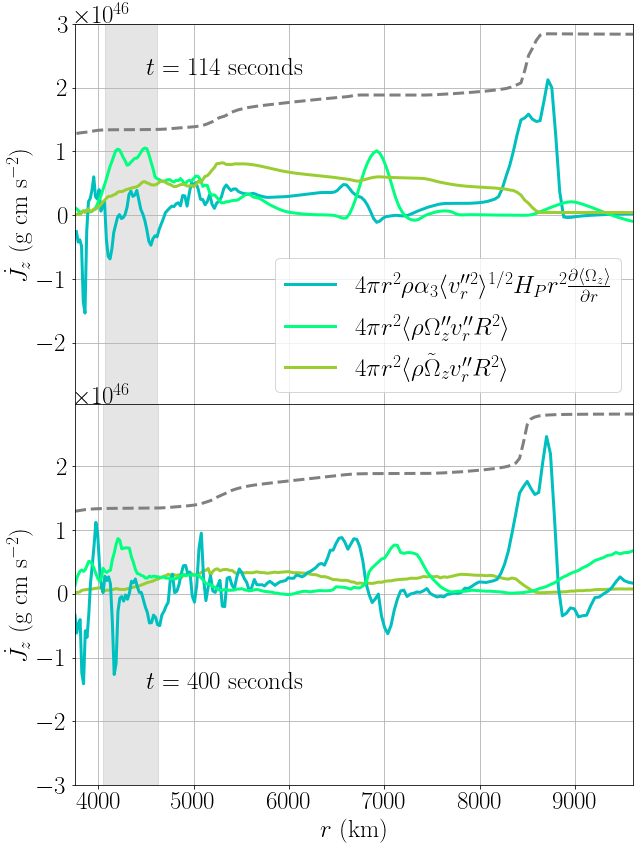}\par 
  \end{multicols}
\caption{Angular momentum fluxes $\dot{J}_z$ at $t = 114\, \mathrm{s}$   after convection is fully developed but before a quasi--equilibrium state is reached
(top panels), and at $t=400\, \mathrm{s}$ during the phase of quasi-stationary convection in the grey-shaded inner convective shell (bottom panels). The structure of the star is indicated by entropy profiles (dashed), which show three convective shells at about $4,000\texttt{-}4,500\, \mathrm{km}$,
$7,000\, \mathrm{km}$, and outside $8,500\, \mathrm{km}$, of which only the inner shell reaches a quasi-stationary state.
The left panels compare the turbulent angular momentum flux from a spherical Favre composition with a base state of constant angular momentum as defined by Equation~(\ref{eq:fturbl})
(light pink) to the diffusive MLT flux expected in case of adjustment to constant specific angular momentum  as defined by
Equation~(\ref{eq:MLTvr}) with $\alpha_3=0.1$ (bright pink).
The right panels compare the turbulent  angular momentum flux (seafoam) and the meridional circulation flux (green) from
Equation~(\ref{eq:momentum_omega}) for a base state of the
flow decomposition with constant $\Omega_z$, to the diffusive MLT flux expected for adjustment to uniform angular velocity with $\alpha_3=0.05$ (teal).
All curves have been obtained using spatial averages over 6 radial zones and $10\, \mathrm{s}$ in time.\\
Even with these small values for $\alpha_3$, the MLT fluxes are qualitatively different from the actual turbulent angular momentum fluxes. In the two inner convective zones, the direction of the turbulent angular momentum flux is clearly opposite to what is required to achieve constant specific angular momentum.
}
\label{fig:MLT-early}
\end{figure*}

The fluxes from Equations~(\ref{eq:MLTvr}) and (\ref{eq:fturbl}) at a time of $114\, \mathrm{s}$
are plotted in Figure~\ref{fig:MLT-early} as hot pink and violet curves using  $\alpha=1$ and $\alpha_3=0.1$.  The mere fact that such an extreme choice of $\alpha_3=0.1$ is required to obtain an MLT flux $F_{\mathrm{MLT},l}$ of the same scale as $F_\mathrm{turb}$ already makes it obvious that MLT does \emph{not} naturally describe the turbulent flux of angular momentum in the oxygen burning shell. By and large, the MLT flux does not even have the same \emph{sign} as the turbulent angular momentum flux. This situation is not unique to the snapshot shown in Figure~\ref{fig:MLT-early} and leads to two conclusions. First, the turbulent angular momentum flux is clearly inconsistent with the assumption that convection homogenises the specific angular momentum and tends to increase rather than decrease the angular momentum gradient in a situation that is still relatively close to uniform rotation. Second, unexpectedly small flux or source terms in a turbulent flow decomposition often suggest that some form of quasi-equilibrium balance has been established. The fact that the turbulent fluxes are so small compared to the expected MLT fluxes could therefore indicate that the differential rotation in the convection zone has already adjusted to some form of quasi-equilibrium.  Further evidence to support this will be discussed later in Section~\ref{section:diff-rotation}.

Next, we consider the MLT flux necessary to drive the model towards a state of rigid rotation. Following the same steps that led to Equation~(\ref{eq:MLTvr}), the MLT flux $F_\mathrm{MLT,\Omega}$ becomes,
	\begin{equation}
	\label{eq:MLTom}
	F_\mathrm{MLT,\Omega}  =4 \pi r^2
	\alpha\, \alpha_3
	\rho \langle v_r''^2\rangle^{1/2}  H_P r^2 \frac{\partial  \Omega_z}
	{\partial r},
	\end{equation}
	which can again be compared to terms in the Favre decomposition of the angular momentum equation. If the base state for the Favre decomposition has constant $\Omega_z$ on shells, an additional term appears in the  Favre-averaged angular momentum equation (see derivation in Appendix~\ref{sec:appendix}), 
	\begin{equation}
    \frac{\pd \langle \rho {l}_z \rangle}{\pd t}
+
\nabla\cdot \mathbf{F}_\mathrm{adv}
+
\nabla\cdot \langle \rho v_r'' 
r^2 \sin^2\theta \rangle \tilde{\Omega}_z
+
\nabla\cdot \langle \rho v_r'' \Omega_z''
r^2 \sin^2\theta
\rangle
=0.
\label{eq:momentum_omega}
\end{equation}
The additional cross term between the spherical average ${\tilde\Omega}_z$
and the fluctuation term $v_r''$ is identical to the meridional
circulation term in \citet{Maeder1998}
and Equation~(\ref{eq:MZ98}), and
can be be expressed in terms of the
quadrupole component $v_{20}$ of the radial velocity component,
\begin{equation}
    \langle \rho r^2 v_r''  \sin^2\theta \rangle \tilde{\Omega}_z= 
    -\frac{4}{3}\sqrt{\frac{\pi}{5}}
    r^2 \bar{\rho} \tilde{v}_{20}   \tilde{\Omega}_z,
\end{equation}
where the multipoles $v_{\ell m}$ are given by
\begin{equation}
    v_{\ell m} = \frac{\int \rho v_r Y^{*}_{\ell m}\, \ud \omega} {\int \rho \, \ud \omega}.
\end{equation}
The putative MLT flux from Equation~(\ref{eq:MLTom}), shown in teal, can again be compared with the two Favre terms from the 3D  simulation  in Figure~\ref{fig:MLT-early}, which are shown in seafoam (turbulent) and yellow (meridional circulation) shades of green. 
The dimensionless coefficients
are set to $\alpha=1$ and $\alpha_3 = 0.05$ for this comparison. Again, an extreme choice of $\alpha_3$ is required to obtain an angular momentum flux that is of comparable magnitude to the actual turbulent fluxes. Even then, the MLT angular momentum flux does not approximate the sum of the turbulent and meridional circulation terms, indicating that the convective zones are not driven towards uniform rotation either, even if the deviation of the angle-averaged angular velocity profiles from uniform rotation in
Figure~\ref{fig:AM-gradient} is modest.

It is noteworthy that the turbulent and meridional circulation terms proportional to $\langle v_r'' \Omega_z''\rangle$ and  $\langle v_r'' \sin^2 \theta \rangle \tilde{\Omega}_z$ are of comparable magnitude, though the meridional circulation term tends to be somewhat smaller inside the convection zones. The sizable contribution of the meridional circulation term indicates that large-scale turnover motions occur and influence the distribution of angular momentum. This is confirmed by further analysis.

\subsection{Rotation as a Function of Latitude}
\label{section:diff-rotation}
\begin{figure*}
\begin{multicols}{2}
\includegraphics[width=\linewidth]{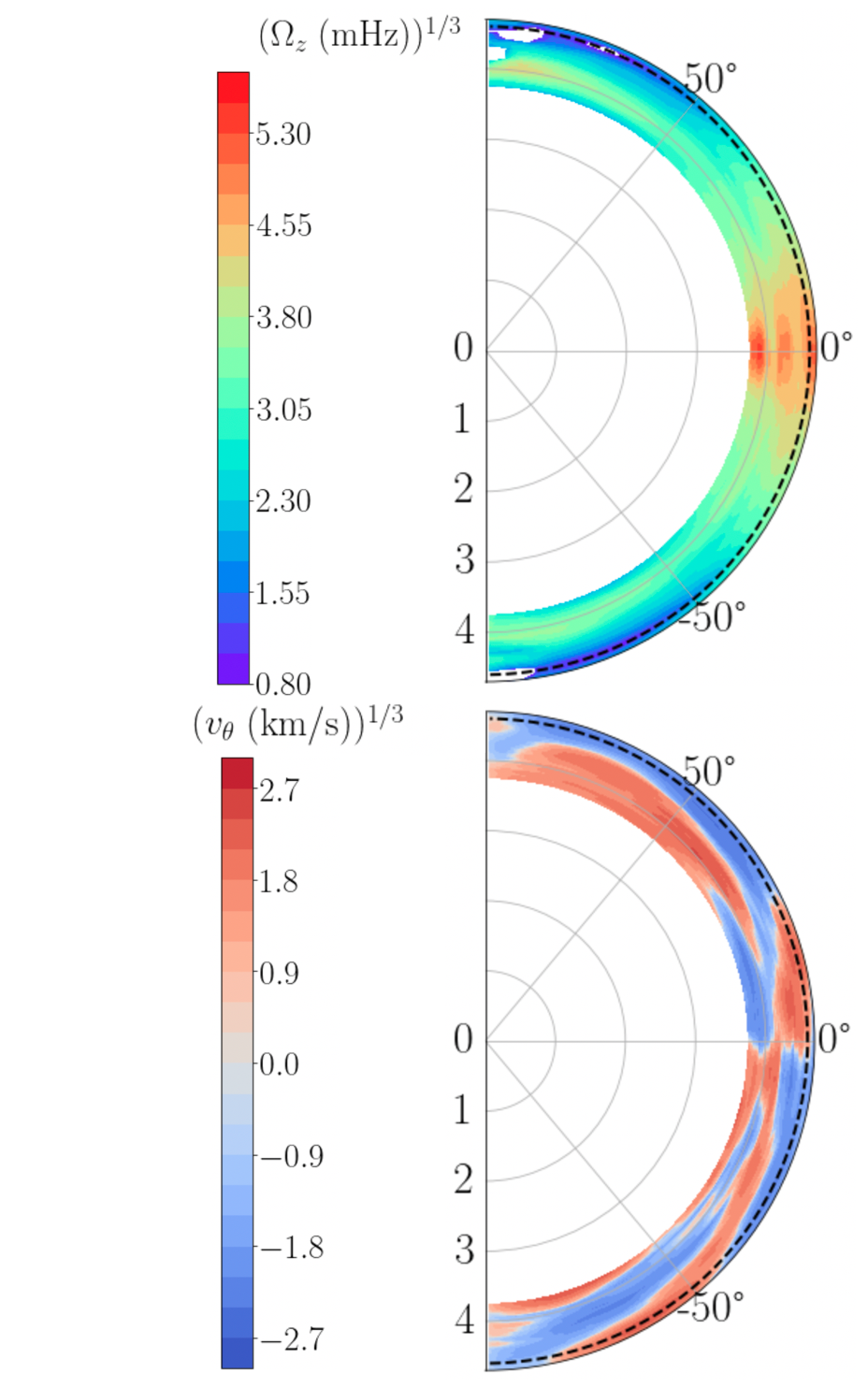}\par 
  \includegraphics[width=\linewidth]{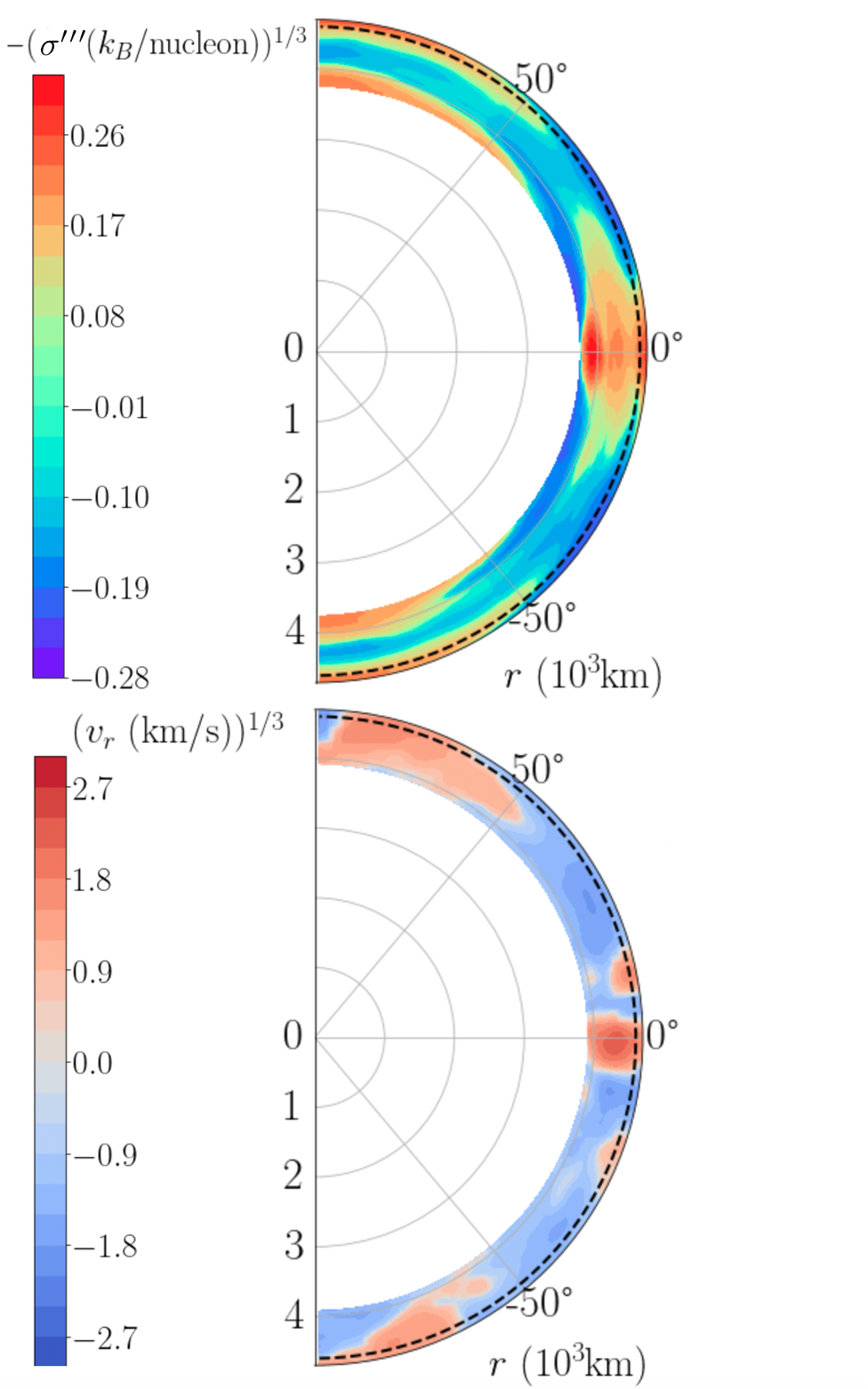}\par 
  \end{multicols}
\caption{2D Favre averages over longitude $\varphi$ of the angular velocity $\Omega_z$
(top left), the meridional component $v_\theta$ of the velocity (bottom left), the radial component $v_r$ of the velocity (bottom right), and the entropy contrast $\sigma''{'}$ (top right) in the inner convective region between about $4000\, \mathrm{km}$ to $4600\, \mathrm{km}$ at $t= 400\, \mathrm{s}$ or 6.7 minutes. 
 Note that we plot the cube root of all quantities in order to enhance the contrast at values close to zero. The outer boundary of this inner convective oxygen shell is shown as a dashed black semicircle. {The convective boundary is defined as the radial coordinate at which the radial gradient of the spherically averaged entropy profile stops being negative.}\\
The angular velocity (top left) clearly shows equatorial superrotation due to the redistribution of angular momentum away from the axis of rotation. Starting from an initial configuration with solid-body rotation at $\Omega_z= 72\, \mathrm{mHz}$ everywhere,
we see a speed-up to  $\Omega_z\approx 100\,\mathrm{mHz}$ at the equator, and slowdown to $\Omega_z \approx 60 \, \mathrm{mHz}$ at mid-latitudes, i.e. rotation at the equator is a few ten{s of} percent faster than at mid-latitudes. The entropy contrast $\sigma'''$ (top right) as defined by Equation~(\ref{eq:ent-def}) exhibits {isocontours similar to} $\Omega_z$, reminiscent of the situation in the solar convection zone \citep{Balbus2009a,Balbus2009b,Balbus2009c}.
The radial and meridional component of the velocity (bottom row) show a meridional circulation pattern with rising flow a the equator, sinking flow at mid-latitudes, and perhaps another circulation cell near the pole.
}
\label{fig:4panel}
\end{figure*}

To further elucidate the differential rotation of the inner convective shell (oxygen burning), we calculate azimuthal averages $\bar{\Omega}_z(r,\theta)$  of the angular velocity, which are plotted in Figure~\ref{fig:4panel} (top left panel) at a time of $400\, \mathrm{s}$. This plot clearly demonstrates that the convective shell (which extends out to the dashed black boundary) neither conforms to the expectations of uniform rotation or uniform specific angular momentum (which would imply that $\Omega_z$ is constant on cylinders and decreases with the distance from the rotation axis). Instead, the shell evolves away from the initial conditions with  rigid rotation at $\Omega_z=72\,\mathrm{mHz}$ to differential rotation  that is fastest at the equator with $\Omega_z\approx 100\, \mathrm{mHz}$. In fact this pattern of differential rotation already emerges  after a few convective turnovers from about $100\, \mathrm{s}$, consistent with  indications for a quasi-equilibrium state from our discussion of Figure~\ref{fig:MLT-early} in Section~\ref{sec:1d_favre}. The latitudinal variation of the angular velocity is much more pronounced than the modest radial dependence  of differential rotation visible in the spherically averaged profiles in Figure~\ref{fig:AM-gradient}.

Figure~\ref{fig:4panel} also shows azimuthal averages of the radial and meridional velocity components $v_r$ and $v_\theta$ (bottom row).  These clearly reveal large-scale meridional circulation with predominant rising flow at the equator, which diverges away from the equator at larger radii, then sinks again at mid-latitudes and flows back to the equator. Another one or two circulation cells can be discerned at higher latitude, even though the third one close to the pole is less conspicuous. 

The pattern of differential rotation is strikingly reminiscent of the Sun \citep{Howe2009} and Sun-like stars \citep{Benomar2018}, even if the quantitative
similarity with $\mathord{\sim} 50 \%$ faster rotation a the equator compared to mid-latitudes
may just be incidental. Intriguingly, the similarities may go further. Closer inspection reveals that our simulation also exhibits a similar close relationship between the rotational isocontours and the azimuthally-averaged entropy contrast  as was noted for the Sun by \citet{Balbus2009b} in the context of their discussion of thermal wind balance.  To illustrate this, we calculate the azimuthally-averaged entropy contrast
$\sigma'''$ (corresponding to $\sigma'$ in
\citealt{Balbus2009b}),
\begin{eqnarray}
   {\sigma'''} (r,\theta) &=& \frac{ \int \rho \sigma''
\, d \varphi}{\int \rho
\, d \varphi}
=\bar{\sigma}-\tilde{\sigma},
\label{eq:ent-def}
\end{eqnarray}
and show the result at $400\, \mathrm{s}$ in
the top right panel of Figure~\ref{fig:4panel}.
We find the same correspondence between
$\sigma'''$ and $\Omega_z$ as in Figure~2 of \citet{Balbus2009b}. Interestingly there is an entropy deficit (negative $\sigma'''$) at the equator (red/orange), which also mirrors the situation in the Sun, suggesting that similar physics is at play. Whether this phenomenology is strictly determined by thermal wind balance cannot be fully ascertained in this work, given that the interpretation of the solar data is not finally settled either \citep[e.g.,][]{Brun2010}. Compared to simulations of the solar convection zone, a detailed verification of force balance is much more complicated because convection never achieves the same degree of quasi-stationarity both because of the limited number of convective turnovers and the \emph{physical} non-stationarity introduced by the contraction of the excised core. Nonetheless, the finding of increased rotational velocity, an entropy deficit, and rising flow at the equator suggests that some form of balance between inertial and buoyancy forces must be at play. To sustain the rising flow at the equator, inertial forces must compensate and slightly outweigh the negative buoyancy force. Longer simulations of earlier, more stationary phases of  late-stage convective burning, perhaps coupled with higher resolution, may be able to better establish how inertial and buoyancy forces and turbulent stresses determine the quasi-equilibrium state of the flow by resorting to Reynolds decomposition with longer temporal averaging than is possible for our simulation.

\begin{figure}
  \centering
    \includegraphics[width=\linewidth]{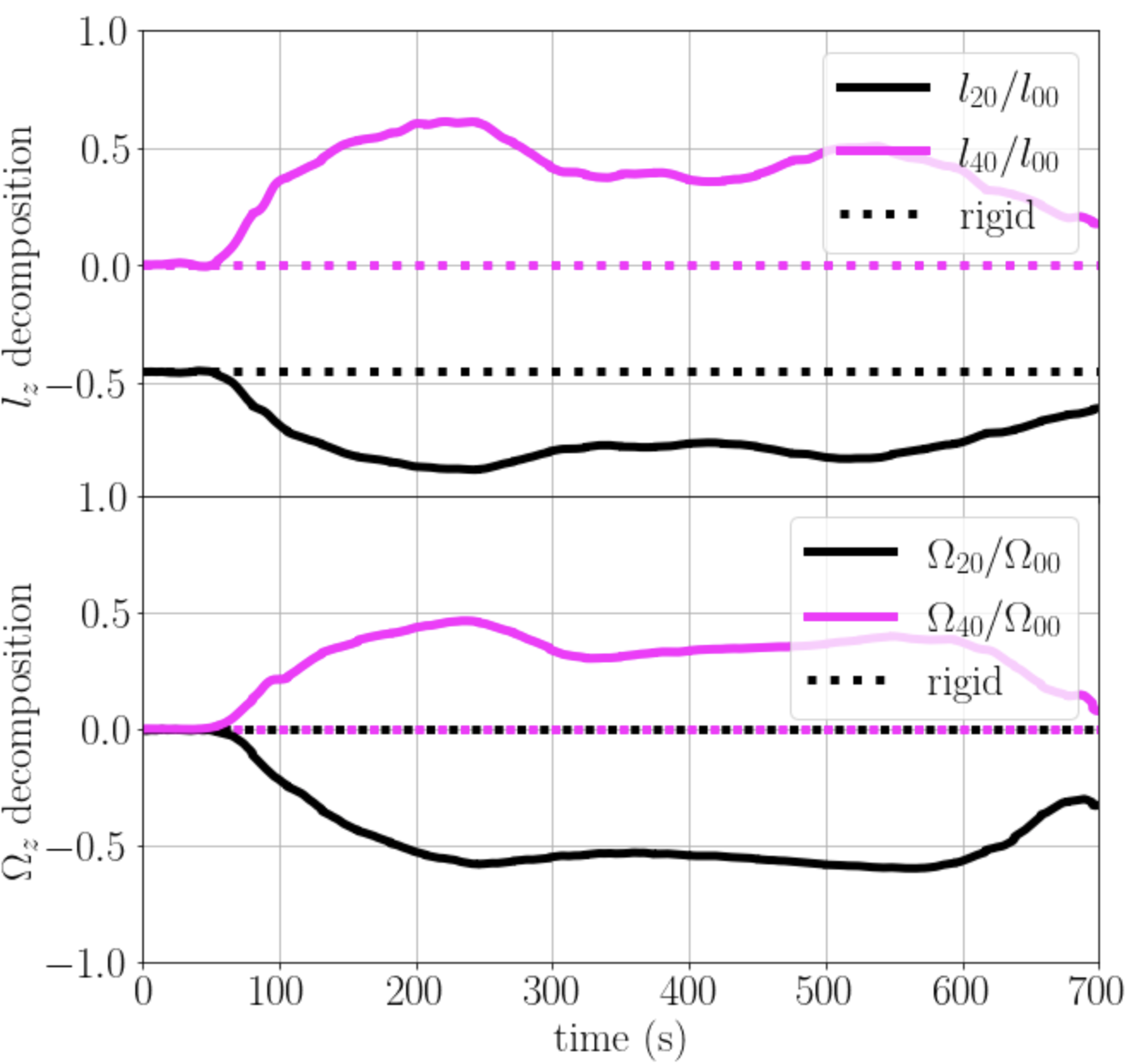}
  \caption{Normalised multipole coefficients
  for the specific angular momentum $l_z$ (top) and  the angular velocity $\Omega_z$ (bottom)
  for the inner convective region over time; see Equations~(\ref{eq:lz3d})
  and (\ref{eq:Om3d}) for definitions.
  Only the lowest-order even multipoles
  $\ell=2$ and $\ell=4$ with axisymmetry ($m=0$) are shown. Initially, $l_{\ell m}$ and $\Omega_{\ell m}$ start at values corresponding to rigid rotation  (dotted lines). 
 During the phase of quasi-stationary convection  
  from $300\, \mathrm{s}$ to $500\, \mathrm{s}$, the normalised quadrupole coefficients reach $l_{20}/l_{00}\approx-0.77$ and $\Omega_{20}/\Omega_{00}\approx-0.45$,   $l_{40}/l_{00}\approx0.37$ and $\Omega_{40}/\Omega_{00}\approx0.35$  indicating superrotation at the equator  (since $Y_{20}$ is negative  and $Y_{40}$ is positive at the equator).  }
\label{fig:Jz-SH}
\end{figure}

\subsection{Quantitative Measures for Differential Rotation}
In order to ascertain that the differential rotation described in the preceding sections has reached some form of quasi-stationary state, it is desirable to consider time-dependent measures of differential rotation.  To this end, we decompose the specific angular momentum $l_z$ and the angular velocity $\Omega_z$ into spherical harmonics $Y_{\ell m}$. To obtain integrated measures of differential rotation across the inner (oxygen burning) shell, we integrate the radius-dependent expansion coefficients  over the entire shell, resulting in the global measures $l_{\ell m}$ and $\Omega_{\ell m}$ for the differential rotation within the shell,
\begin{align}
l_{\ell m} &= \iint  Y_{\ell m}^*(\theta,\varphi) \, l_z(r,\theta,\varphi) \,\ud \omega\,r^2 \ud r
\\
&
=
 \iint  Y_{\ell m}^*(\theta,\varphi) \, r^2 \Omega_z(r,\theta,\varphi) \sin^2 \,\ud \omega\, r^2 \ud r
\label{eq:lz3d}
\\
\Omega_{\ell m} &= \iint  Y_{\ell m}^*(\theta,\varphi) \, \Omega_z(r,\theta,\varphi) \,\ud \omega\, r^2 \ud r
\label{eq:Om3d}.
\end{align}
The integrals run over solid angle $\omega$ and the (time-dependent) radial bounds in $r$ which define the convective burning region via the spherically averaged entropy profile (e.g., Figure~\ref{fig:composition}). The differential rotation is most conveniently characterised by the normalised coefficients
$\Omega_{\ell m}/\Omega_{00}$ and $l_{\ell m}/l_{00}$, which also mitigates artificial trends in  $\Omega_{\ell m}$ and $l_{\ell m}$ that may be due to numerical spin-down.
It is useful to note that for the special case of rigid rotation, $\Omega_{00}=1$,
$l_{00}=1$ and $l_{20}={-}1/\sqrt{5}$ (in the limit of thin shells) are the only non-vanishing coefficients.\footnote{This can be derived by noting that
$l_z=r^2\sin^2 \theta \Omega_z = r^2 \left( \frac{4}{3} \sqrt{\pi}Y_{00} -\frac{4}{3} \sqrt{\frac{\pi}{5}} Y_{20}\right)\Omega_z $, so for $\Omega_z=\mathrm{const.}$ only the coefficients with $\ell=0$ or $\ell=2$ and $m=0$ remain in the spherical harmonics expansion of the specific angular momentum.
} For the case of constant specific angular momentum, all coefficients $l_{\ell m}$ except
$l_{00}$ would be zero, and the coefficients
$\Omega_{\ell m}$ will tend to $\Omega_{\ell m}/\Omega_{00}=Y_{\ell m}/Y_{00}$ for even $\ell$
and $m=0$ and vanish otherwise.\footnote{For $l_z=\mathrm{const}.$, one has $\Omega_z=l_z/(r^2 \sin^2 \theta)$, and the
ratio  $\Omega_{\ell m}/\Omega_{00}$ is determined by the values of $Y_{\ell m}^*$ at the poles of the integrand in Equation~(\ref{eq:Om3d}).}

To characterise the large-scale patterns in the differential rotation previously shown in Figure~\ref{fig:4panel}, we focus on the lowest non-trivial multipoles $\ell=2$
and $\ell=4$ with equatorial symmetry and zonal wavenumber $m=0$, i.e., on the  coefficients
$\Omega_{20}/\Omega_{00}$,  $\Omega_{40}/\Omega_{00}$, $l_{20}/l_{00}$, and  $l_{40}/l_{00}$. These coefficients are plotted in Figure~\ref{fig:Jz-SH}, and their values in the simulation and for the limiting cases of constant angular velocity and constant specific angular momentum are summarised in Table~\ref{tab:coefficients}.

Once convection becomes non-linear after about
$50\, \mathrm{s}$, the coefficients $\Omega_{20}/\Omega_{00}$, $\Omega_{40}/\Omega_{00}$,
and $l_{40}/l_{00}$ start to deviate from zero, and $l_{20}/l_{00}$ takes on larger negative value compared to its initial state of
$l_{20}/l_{00}\approx -0.45$. Consistent with the evolution \emph{away} from constant specific angular momentum $\Omega_{20}/\Omega_{00}$ becomes negative rather than positive.

\begin{table}
\caption{Lowest-order even multipole coefficients of the spherical harmonics decomposition of angular momentum and angular velocity for the simulation and the limiting cases of constant angular momentum and constant angular velocity.
Most of these values are included in Figure~\ref{fig:Jz-SH}.}
\label{tab:coefficients}
\begin{center}
\begin{tabular}{|c|c|c|c|}
\hline
\textbf{Rotation} & \textbf{Multipole} & \textbf{Normalised } & \textbf{{Line type} in}\\
\textbf{pattern} & & \textbf{value} & \textbf{Figure~\ref{fig:Jz-SH}} \\
& & & \textbf{(t)op/(b)ottom}
\\ \hline
Rigid & $l_{20}$ &  -0.45 & dotted black (t) \\ \hline
Rigid & $l_{40}$ &  0 & dotted magenta (t) \\ \hline
Rigid & $\Omega_{20}$ &  0 & dotted black (b) \\ \hline
Rigid & $\Omega_{40}$ &  0 & dotted magenta (b)\\ \hline
Constant $l_z$ & $l_{20}$ &  0  & not shown  \\ \hline
Constant $l_z$ & $l_{40}$ &  0 & not shown \\ \hline
Constant $l_z$ & $\Omega_{20}$ &  2.24 & not shown \\ \hline
Constant $l_z$ & $\Omega_{40}$ &  3 & not shown \\ \hline
Simulation & $l_{20}$  & -0.77  & solid black (t) \\ \hline
Simulation & $l_{40}$  & 0.37  & solid magenta (t) \\ \hline
Simulation & $\Omega_{20}$  & -0.55  & solid black (b) \\ \hline
Simulation & $\Omega_{40}$  & 0.35  & solid magenta (b) \\ \hline
\end{tabular}
\end{center}
\end{table}

After about $200\, \mathrm{s}$, we see indications of a quasi-stationary state, as the normalised multipole coefficients plateau and then fluctuate around their long-term averages without any discernible secular trend for several hundred seconds. During the period between $300\, \mathrm{s}$ and $500\, \mathrm{s}$, we find average values of about
$\Omega_{20}/\Omega_{00}=-0.55$, $\Omega_{40}/\Omega_{00}=0.35$,
 $l_{20}/l_{00}=-0.77$, and $l_{40}/l_{00} = 0.37$. Only shortly before the onset of collapse do the normalised multipole coefficients diminish. It is likely that at this stage the adjustment to a quasi-stationary state is no longer fast enough to keep pace with the contraction of the shell and its concomitant spin-up by angular momentum conservation.

\begin{figure}
  \centering
    \includegraphics[width=0.5\textwidth]{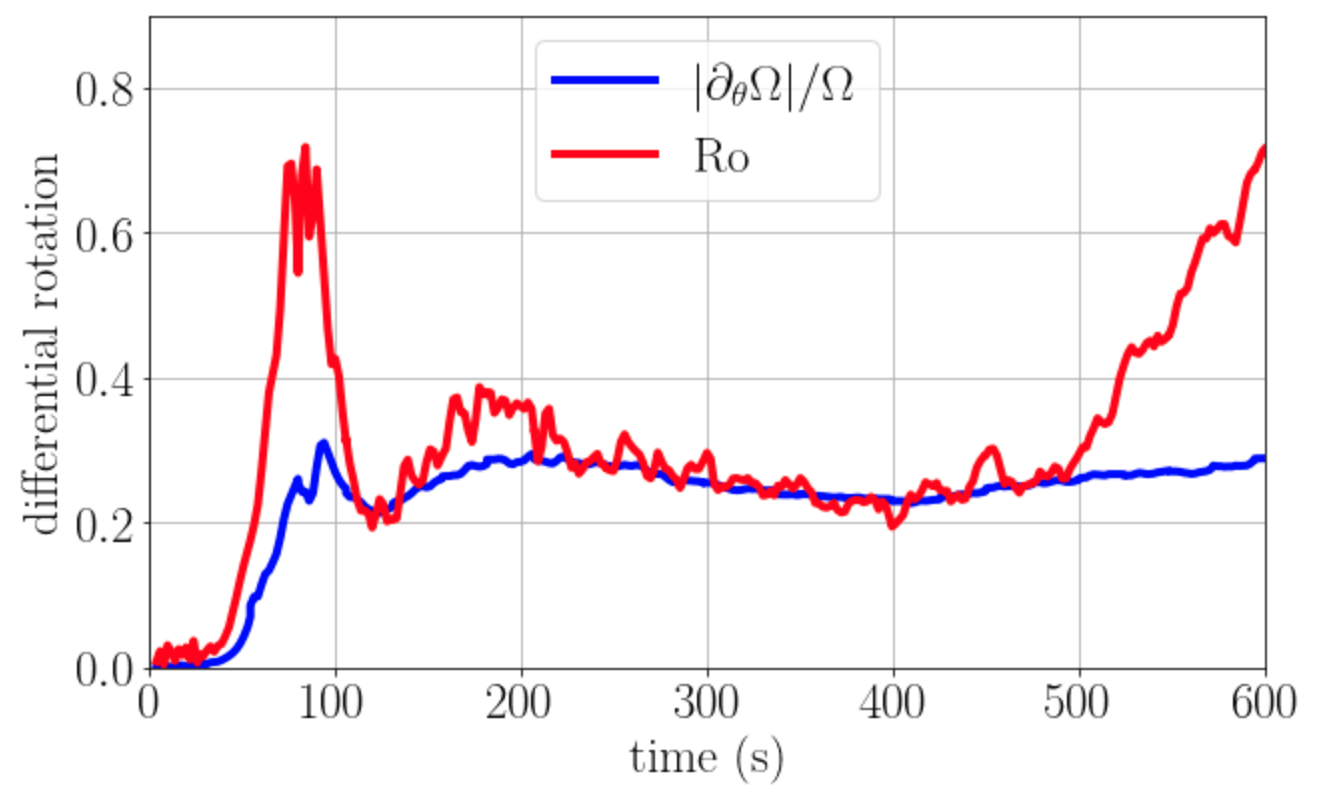}
  \caption{The global measure  of differential rotation,
  $|\partial_\theta \Omega|/\Omega$ as defined by Equation~(\ref{eq:diff-rotation}) (blue curve)
  and the convective Rossby number in the oxygen shell (red curve) as a function of time, shown until $600\, \mathrm{s}$. The convective shell is in the regime of moderately fast rotation where the scaling laws of \citet{Jermyn2020a,Jermyn2020b} predict a scaling
  $|\partial_\theta \Omega|/\Omega\sim \mathrm{Ro}^{3/5}$. The degree of differential rotation in the oxygen shell is roughly consistent with this prediction.
  }
\label{fig:diff-rotation}
\end{figure}

To conclude our analysis of the shell burning model, we note that the degree of differential rotation roughly fits recently proposed analytic scaling relations from \citep{Jermyn2020a} and trends in simulations and observations of rotating convection zones in other contexts, e.g., surface convection zones, as recently summarised by \citet{Jermyn2020b}. For this purpose, another measure for differential rotation is required.
Among the various measures for differential rotation in the literature, we opt for a metric similar to \citet{Brown2008}, which is
one of the studies included in the comparison of theory, simulations, and observations by \citet{Jermyn2020b}. We define an overall measure  $|\pd_\theta \Omega|/\Omega$ for differential rotation in latitude based on the root-mean-square deviation of the angular velocity from solid-body rotation at a given radius,
\begin{equation}
\label{eq:diff-rotation}
     \frac{|\partial_\theta \Omega|}{\Omega} = \frac{ (\int \langle  \Omega''(r,\varphi,\theta)^2 \rangle \, \ud V)^{1/2}}{\int \tilde{\Omega}_z(r) \, \ud V },
\end{equation}
where $\tilde{\Omega}_z(r)$ is the spherically-averaged angular velocity, and where the domain of integration is the oxygen burning shell.
 We plot this measure of differential rotation as a function of time in Figure~\ref{fig:diff-rotation} (blue) and find that
$|\pd_\theta \Omega|/\Omega$ is
very stable once convection has fully developed around $100\, \mathrm{s}$ , maintaining values of $\mathord{\approx} 0.28$ during the quasi-stationary period between $300\, \mathrm{s}$ and $500\, \mathrm{s}$. This again confirms that a quasi-stationary angular momentum distribution has been reached.

The theoretical analysis of \citet{Jermyn2020a} suggests (extending
earlier ideas, e.g., by \citealt{Showman2011})
that the degree of differential rotation should be determined by the convective Rossby number $\mathrm{Ro}$, which appears to roughly fit the evidence from simulations and observations \citep{Jermyn2020b}.
\citet{Jermyn2020a,Jermyn2020b} predict
$|\pd_\theta \Omega|/\Omega\sim 1$ in the regime of slow rotation ($\mathrm{Ro}>1$)
and a power-law dependence for fast rotation
($\mathrm{Ro}<1$),
\begin{equation}
    \frac{|\partial_\theta \Omega|}{\Omega} \sim \mathrm{Ro}^{3/5}.
\end{equation}
{Here we note that these scalings do not discern between solar (this simulation) and anti-solar \citep{Arnett2010,Yoshida2021} differential rotation patterns}. \citet{Jermyn2020a} define the convective Rossby as $\mathrm{Ro}=N_\mathrm{BV}/\Omega$ using the Brunt-V\"ais\"al\"a frequency $N_\mathrm{BV}$; in \citet{Jermyn2020b} prefactors different from one are sometimes used. Since it is difficult to measure the superadiabatic gradient in 3D simulations to obtain $N_\mathrm{BV}$, we rather compute
$\mathrm{Ro}$ based on the convective velocity, similar to \citet{Brun2005},
\begin{equation}
    \mathrm{Ro} = \frac{v_\mathrm{conv,O}}{2 \Omega_\mathrm{shell} \,\Delta r} = \frac{1}{2\tau_\mathrm{conv} \Omega_\mathrm{shell}}.
    \label{eq:Rossby}
\end{equation}
Here $v_\mathrm{conv,O}$ is the average convective velocity in the shell, defined by the maximum
root-mean-square value of 
$\langle v_r''^2 \rangle^{1/2}$ of the Favrian radial velocity fluctuations in the convective zone, $\Delta r$ is the width of the oxygen burning shell,  $\tau_\mathrm{conv} = \Delta r/v_\mathrm{conv,O}$ is the convective turnover time, and $\Omega_\mathrm{shell}$ is the shell-averaged angular velocity. {The factor 2 in Equation~(\ref{eq:Rossby}) comes from the Coriolis force $2 \boldsymbol{\Omega} \times \mathbf{v}$ \citep[e.g.][]{goldstein}, which translates into a factor 2 in the Coriolis frequency $\Omega_\mathrm{C} $ that actually enters in the usual definition of the Rossby number}.
Considering the order-of-magnitude nature of the proposed scaling laws of \citet{Jermyn2020a}, ambiguities that might introduce small factors of order unity in the definition of the Rossby number are immaterial, as is a possible mild deviation between the convective turnover frequency $1/\tau_\mathrm{conv}$ and the 
Brunt-V\"ais\"al\"a frequency $N_\mathrm{BV}$. We find that the convective Rossby number is of order  $\mathrm{Ro}\approx 0.25\texttt{-}0.4$  during the phase of quasi-stationary convection (Figure~\ref{fig:diff-rotation}, red curve), placing it in the regime of moderately fast rotation. The mild degree of differential rotation is therefore broadly consistent (within a factor of $\mathord{\sim} 2$) with the scaling relation of \citet{Jermyn2020a} and other  compressible 3D hydro simulations of surface convection at similar Rossby number by \citet{Kapyla2011,Mabuchi2015}, 
see Figure~6 of \citet{Jermyn2020b}.

{For comparison, we also estimate $\mathrm{Ro}$ for the \cite{Yoshida2021} simulation. Using their averaged values of the turbulent velocity and the rotational
velocity in the Si/O-rich layer ($5.5 \times 10^7 \,\mathrm{cm}\,\mathrm{s}^{-1}$ and $2.0 \times 10^7\,\mathrm{cm}\,\mathrm{s}^{-1}$ respectively), we estimate their convective Rossby number to be $\mathrm{Ro} = 2.75$, i.e. in the regime of ``slow rotation'' ($\mathrm{Ro}>1$). If these global scalings are correct, and the simulation has reached a steady state, then one expects an even more extreme level of differential rotation in the latitudinal direction, compared to the simulation presented here.}

\section{Conclusion{s}}
\label{sec:conlclusions}
We have presented a study of differential rotation in
 the convective oxygen burning shell of a rapidly rotating $16\, M_\odot$ helium star based on a 3D simulation with the \textsc{Prometheus} code 
\citep{fryxell_89,fryxell_91}. Starting from a {1D} stellar evolution model with rigid rotation in convection zones,
we find that the convective oxygen shell naturally develops differential rotation, with the angular momentum distribution reaching a quasi-steady state after a few turnover times. Different from previous simulations in 2D \citep{Arnett2010} and 3D \citep{Yoshida2021}, the rotation profile does not tend towards uniform specific angular momentum in the convective zone. Instead the differential rotation is characterised by a positive angular velocity gradient and faster rotation at the equator by several tens of percent.

We performed a turbulent flow decomposition based on
spherical and azimuthal Favre averages to quantitatively analyse the distribution and transport of angular momentum within the computational domain. The flow decomposition shows that the turbulent transport of angular momentum is clearly not described by a 1D mixing-length theory ansatz with diffusive fluxes that drive the rotation profile towards constant angular velocity or constant specific angular momentum. {Because of significant fluctuations in the spherically averaged angular momentum and angular velocity profiles, it is, however, more difficult to judge the magnitude of the radial angular velocity gradient and the significance of the deviation from uniform rotation in an effective 1D description as used in stellar evolution models. There is much stronger evidence for differential rotation with latitude.} 2D flow averages reveal a global flow pattern with meridional circulation, excess angular velocity, and an entropy deficit at the equator. A quantitative analysis shows that the degree of differential rotation remains stable for several hundred seconds after an initial adjustment phase, suggesting that a quasi-equilibrium state has been reached. When the oxygen shell contracts significantly and the burning speeds up shortly before collapse, the quasi-equilibrium can no longer be maintained, and the ``superrotation'' of the equatorial region becomes less pronounced. 

The pattern of differential rotation found in our simulation is conspicuously similar to observations of the Sun \citep{Howe2009} and Sun-like stars \citep{Benomar2018}. Our model roughly fits recently proposed analytic scaling laws that relate the degree of differential rotation to the convective Rossby number \citep{Jermyn2020a,Jermyn2020b}, and is consistent with compressible simulations of stellar surface convection at similar Rossby numbers of $\mathord{\sim}0.2\texttt{-}0.4$ \citep[e.g.,][]{Kapyla2011,Mabuchi2015}.  This suggests that similar mechanisms may be at play as in surface convection zones. Specifically, the meridional flow pattern with superrotation and an entropy deficit at the equator may be explained as a result of thermal wind balance \citep{Balbus2009b}. Due to the more dynamical nature of convective burning in our model compared to solar surface convection, a detailed analysis of angular momentum transport and force balance within the convective zone is not feasible within the current study. Longer simulations of more steady phases of late-stage burning may reveal more clearly how buoyancy, inertial forces, and turbulent stresses determine the pattern of differential rotation. Future simulations should also include magnetic fields, although these may not change the picture qualitatively based on evidence from observations and simulations
\citep[for an overview see][]{Jermyn2020b}.
It is also desirable to confirm the current results at higher resolution and minimise numerical non-conservation of angular momentum, even though we have used a formulation of the momentum equation that improves angular momentum conservation \citep{Mueller2020}. However, our analysis of turbulent fluxes and the large-scale meridional flow structure suggests that the superrotation that we observe is physical and not due to a numerical violation of angular momentum conservation.

Since our findings are clearly different from the 2D simulation of \citet{Arnett2010} and the recent short 3D simulation of \citet{Yoshida2021}, further investigation of late-stage convection in rotating massive stars is required. There is evidently a tension between their results and ours, as our study suggests that convection may steepen rather than flatten specific angular momentum gradients. However, there is not necessarily a contradiction. The 2D results of \citet{Arnett2009} may well be due to the constraint of axisymmetry, which can qualitatively affect the nature of turbulent convective flow.  The 3D results of \citet{Yoshida2021} only covered few convective turnovers and may not have reached steady-state differential rotation. During an initial dynamical adjustment phase, it is conceivable that advective transport may effect some homogenisation of specific angular momentum before a true steady state is reached. Such a phenomenon may play a role towards the end of our simulation as well when convection speeds up and the equatorial superrotation becomes less pronounced. Given the {conspicuous    similarities} with simulations of stellar surface convection, we believe, however, that our longer simulation probably gives a better representation of the differential rotation that will emerge in the long run and be maintained over evolutionary time scales.
{An alternative explanation could be
that the model of \citet{Yoshida2021} selects an anti-solar pattern of differential rotation. This might make their model conform with the scaling laws of \citep{Jermyn2020a,Jermyn2020b}, which would predict stronger differential rotation for their model, which is characterised by a higher Rossby number than ours. No firm conclusions on the reasons for the disagreement between our simulation and theirs can be drawn at this stage.
}

If the interior convection zones in massive stars are generically characterised by differential rotation with a positive angular velocity gradient, this could have interesting implications for the pre-collapse rotation rates and magnetic fields of supernova progenitors, and in particular for the proposed evolutionary channels for hypernovae and GRB supernovae. The cores of supernova progenitors might be spun down even more strongly than predicted by current stellar evolution models, making it more difficult to achieve the rotation rates required for the millisecond magnetar scenario, but easier to achieve sufficient angular momentum in shells a few solar masses away from the core as required for disk formation in the collapsar scenario \citep{Woosley1993,MacFadyen1999}.
{However, this remains in the realm of speculation at this point and} will need to be investigated using stellar evolution models that appropriately account for superrotation in convective zones. While it is relatively easy to relax the assumption of uniform rotation in convective zones in stellar evolution models
(Heger \& M\"uller in prep.), capturing the nuances of the underlying physics is likely more difficult, especially when the findings from 3D models are not fully understood either. An advection-diffusion approach to angular momentum transport of \citet{Maeder1998} or generalisations thereof \citep{Takahashi2021}
may be an appropriate starting point for projecting the results of 3D simulations into 1D stellar evolution models in a more consistent manner.

\section*{Acknowledgements}
We thank A.~Heger for fruitful discussions and for providing data for progenitor model HE16O during the pre-collapse phase, and I.~Mandel for fruitful discussions. {We thank our referee, R.~Hirschi, for his careful reading of the earlier version, and for helpful suggestions and comments that improved our manuscript.}
LM acknowledges support by an Australian Government Research Training (RTP) Scholarship, and a Monash University Post Graduate Publication Award (PPA). BM has been supported, in part, by the Australian Research Council through a Future Fellowship (FT160100035). This research was undertaken with the assistance of
resources obtained through the National Computational Merit Allocation Scheme and ASTAC from the National Computational Infrastructure
(NCI), which is supported by the Australian Government and
was supported by resources provided by the Pawsey Supercomputing
Centre with funding from the Australian Government and the Government of Western Australia.

\section*{Data availability}
The data underlying this manuscript will be shared on reasonable request to the authors, subject to considerations of intellectual property law. Additional movies can be found \href{https://github.com/mcneilllucy/3D-stellar-hydrodynamics/tree/master/he16}{here}.

\bibliographystyle{mnras}
\bibliography{refs}       



\appendix

\section{Favre-averaged Angular Momentum Equation with Different Base States} 
\label{sec:appendix}
The evolution equation for the specific angular momentum $\mathbf{l}$ can be obtained by taking the cross product of the position vector $\mathbf{r}$ with the momentum equation \citep{shu,Pope2000}, leading to \citep[{e.g.}\ ][]{MM2020},
\begin{equation}
 \frac{\partial \rho \mathbf{l}}{\partial t} 
+  \nabla \cdot (\rho \mathbf{v}\otimes \mathbf{l})
+\mathbf{r} \times \nabla P 
=
\mathbf{r}\times \rho \mathbf{g}.
\end{equation}
Without loss of generality, we can identify the overall axis of rotation with the $z$-axis. The equation for the angular momentum component $l_z$ becomes.
\begin{equation}
 \frac{\partial \rho l_z}{\partial t} 
+  \nabla \cdot (\rho \mathbf{v} l_z)
+\mathbf{e}_z\cdot (\mathbf{r} \times \nabla P)
=
\mathbf{e}_z\cdot (\mathbf{r}\times \rho \mathbf{g}).
\end{equation}
The right-hand side vanishes as long as we assume monopole gravity. The pressure term vanishes if we discard components of $\nabla P$ in the zonal direction ($\varphi$-direction), though not necessarily in the meridional direction
($\theta$-direction), which is a generalisation of the assumption of vanishing horizontal pressure gradients in turbulence models for non-rotating stars \citep{kuhfuss_86,BM2019b}. Integrating over spherical shells then yields
\begin{equation}
 \frac{\partial \int \rho l_z\,\ud \omega}{\partial t} 
+  \nabla_r
\int \rho v_r l_z\,\ud \omega
=
0,
\end{equation}
where $\nabla_r$ is the radial component of the divergence operator. If we perform a Favre decomposition \citep{Favre1965} with a base state $\hat{\rho}$, $\tilde{v}_r$, and $\tilde{l}_z$, we obtain
\begin{align}
 \frac{\partial \langle \rho l_z\rangle}{\partial t} 
+  \nabla_r
\langle \rho (\tilde{v}_r+v_r'') (\tilde{l}_z+l_z'')\rangle&=0,\\
\frac{\partial \hat{\rho}\tilde{l}_z}{\partial t} 
+  \nabla_r
\rho \tilde{v}_r \tilde{l}_z 
+\nabla_r
\langle v_r'' l_z''\rangle
&=
0,
\end{align}
where we have applied the usual rules for Favre averages for  quantities $X$ and $Y$ that are averaged with mass weighting,
\begin{align}
\langle\rho X\rangle &=\hat{\rho} \tilde{X}\\
\langle\rho \tilde{X} \tilde{Y}\rangle&=\hat\rho \tilde{X} \tilde{Y}\\
\langle \rho \tilde{X} Y''\rangle&=0.
\end{align}
If, however, we expand around a base state with constant angular velocity $\tilde{\Omega}_z$ on spheres,
\begin{equation}
\tilde{\Omega}_z=\frac{\langle \rho \Omega_z r^2 \sin^2\theta\rangle}{\langle \rho r^2 \sin^2\theta\rangle} =
\frac{\tilde{l}_z}{\tilde{i}_{zz}},
\end{equation}
with $\tilde{i}_{zz}=\langle \rho r^2 \sin^2\theta\rangle/\hat{\rho}$,
then these rules do not hold any longer. 
After writing the angular momentum equation
in terms of the base state and the fluctuations
\begin{equation}
 \frac{\partial \hat{\rho}\tilde{\Omega}_z \tilde{i}_{zz}}{\partial t} 
+  \nabla_r
\langle \rho (\tilde{v}_r+v_r'')
(\tilde{\Omega}_z+\Omega_z'') r^2 \sin^2\theta\rangle =
0,
\end{equation}
partly expanding the products inside the angle product results in
\begin{align}
\frac{\partial \hat{\rho}\tilde{\Omega}_z \tilde{i}_{zz}}{\partial t} 
+  \nabla_r
&
\Big[
\langle \rho 
\tilde{v}_r (\tilde{\Omega}_z+ \Omega_z'') r^2 \sin^2\theta
\rangle
+\langle\rho v_r'' \tilde{\Omega}_z r^2 \sin^2\theta
\rangle
\nonumber
\\
&
+\langle \rho v_r''\Omega_z'' r^2 \sin^2\theta\rangle
\Big]=
0.
\end{align}
The first term in square brackets can be factored into $\hat{\rho} \tilde{v}_r \tilde{\Omega}_z \tilde{i}_{zz}$, and $\tilde{\Omega}_z$ can be taken out of the angled brackets in the second term. Hence the Favre-averaged angular momentum equation for  a base state with constant angular velocity on shells becomes
\begin{align}
    \frac{\pd \hat{\rho}\tilde{\Omega}_z \tilde{i}_{zz}}{\pd t}
&+
\frac{1}{r^2}\frac{\pd}{\pd r}
\left( r^2 \hat{\rho} 
\tilde{v}_r \tilde{\Omega}_z \tilde{i}_{zz}\right)
+
\frac{1}{r^2}\frac{\pd}{\pd r}
\left(r^2 \tilde{\Omega}_z
\langle \rho v_r'' 
r^2 \sin^2\theta \rangle
\right)
\nonumber
\\
&+
\frac{1}{r^2}\frac{\pd}{\pd r}
\left( \rho r^4
\langle
v_r'' \Omega_z''
 \sin^2\theta\rangle
\right)
=0.
\end{align}

\bsp	
\label{lastpage}
\end{document}